%% file: main.tex
\documentclass[conference]{IEEEtran}
\usepackage{blindtext, graphicx}
\usepackage{amsfonts}
\usepackage{cite}

\ifCLASSINFOpdf
\else
\fi
\usepackage[cmex10]{amsmath}
\usepackage{algorithmic}
\usepackage[tight,footnotesize]{subfigure}
\begin{document}
%
\title{A Novel PMU Fog based Early Anomaly Detection for an Efficient Wide Area PMU Network}

\author{\IEEEauthorblockN{Zekun Yang, Ning Chen, Yu Chen, Ning Zhou}
	\IEEEauthorblockA{Dept. of Electrical and Computing Engineering, Binghamton University, SUNY, Binghamton, NY 13902}
}


%


\maketitle

\input{IEEEfog_1.tex}

\ifCLASSOPTIONcaptionsoff
  \newpage
\fi



%

\bibliographystyle{IEEEtranS}

\bibliography{./ref}

%





\end{document}

%% file: IEEEfog_1.tex
\begin{abstract}
Based on phasor measurement units (PMUs), a synchronphasor system is widely recognized as a promising smart grid measurement system. It is able to provide high-frequency, high-accuracy phasor measurements sampling for Wide Area Monitoring and Control (WAMC) applications. However, the high sampling frequency of measurement data under strict latency constraints introduces new challenges for real time communication. It would be very helpful if the collected data can be prioritized according to its importance such that the existing quality of service (QoS) mechanisms in the communication networks can be leveraged. To achieve this goal, certain anomaly detection functions should be conducted by the PMUs. Inspired by the recent emerging edge-fog-cloud computing hierarchical architecture, which allows computing tasks to be conducted at the network edge, a novel PMU fog is proposed in this paper. Two anomaly detection approaches, Singular Spectrum Analysis (SSA) and K-Nearest Neighbors (KNN), are evaluated in the PMU fog using the IEEE 16-machine 68-bus system. The simulation experiments based on Riverbed Modeler demonstrate that the proposed PMU fog can effectively reduce the data flow end-to-end (ETE) delay without sacrificing data completeness.   
\end{abstract}

\begin{IEEEkeywords}
Phasor Measurement Unit (PMU), Communication Network, Fog Computing, Anomaly Detection, Smart Grid.
\end{IEEEkeywords}

\section{Introduction} 
Nowadays, the quick growing demand and high dependence on electricity, from government,  commercial and household users, are necessitating a high quality power supply. The operators are expected to have a global view of the health condition of the power grid. As a result, there is an increasing interest for wide area monitoring and control (WAMC) in the power grid \cite{delivery2013synchrophasor, fang2012smart}. The synchrophasor system based on phasor measurement unit (PMU) is considered as the most promising measurement technology to support WAMC applications in the future \cite{delivery2013synchrophasor}. The synchrophasor system, also named as the synchronized phasor measurement system, aims to enhance the wide-area monitoring, protection and control system of power grids by providing high accuracy, high sampling rate and synchronized system measurement data. 

The sample rate of the widely used Supervisory Control and Data Acquisition (SCADA) system is one sample every 2-4 seconds. In contrast, the data acquisition devices, PMUs, measure and report the instantaneous voltage phasor, current phasor, and frequency with the speed of 30 to 120 observations per second~\cite{patel2010real}. Meanwhile, the monitoring and assessment applications based on PMUs, including power system fault detection and identification, are designed to exploit real time measurements and have a strict latency requirement~\cite{kansal2012bandwidth}. The delay in the communication networks is a critical factor to meet the real-time delivery requirement for the large amount of data. 

Considering the volume of monitoring data generated by PMUs, it is non-trivial to provide high quality of service (QoS) in such a big data transfer service. One observation is, however, not all the data is of same importance. When the power grid functions smoothly, longer delays are tolerable. What the operators really need is the immediate delivery of data when anomalies happen. Therefore, the capability of prioritizing the monitoring data will enable us to leverage the QoS features of the communication networks, such that the important data receives high QoS including the guaranteed ceiling of delay, average and peak data rate, limited jitters, etc. Now, it is a chicken-egg problem: the operator needs the data reflecting anomalous behaviors as early as possible, but before being processed we cannot tell which data is the useful one.

Thanks to the fast growth of computing technologies, edge-fog-cloud computing paradigm provides us a key to break the chicken-egg dilemma. Edge computing and fog computing extend the realm of cloud computing by allowing more computing intensive tasks at the edge of the network~\cite{shi2016edge}. In the edge-fog-cloud hierarchical architecture, fog computing is attracting more and more attention in a variety of applications \cite{chen2017smart}. On one hand, different from remote cloud centers, fog computing consists of heterogeneous smart devices with close proximity to end users. On the other hand, fog layer possesses more computing resources than the embedded Internet of Things (IoTs) devices at the edge. In addition, as tons of monitoring data are processed continuously at the fog, workload placed on communication networks is reduced significantly.

These advantages make fog computing an ideal solution for the delay sensitive synchrophasor system to improve the efficiency of the PMU communication network and provide high quality data to the WAMC applications. The PMU devices, working with sensing and phasor extraction functions at the substation domain, are scalable and capable for extra computing task. Although it is difficult to make sophisticated controlling decisions with local information, PMUs are able to conduct some data pre-processing work on-site, such as data compressing, feature extraction and so on. In this paper, a novel PMU fog architecture is proposed, which leverages the computing resource in PMU devices to detect and mark anomalous data on-site. The marked data will be transmitted with higher priority in the communication network. Such that the critical data can be received by the controller in a timely manner without sacrificing data quality.  
 
The rest of the paper is organized as follows. The Section \ref{sec:rel} introduces the background and related work on synchrophasor system and fog computing paradigm. Section \ref{sec:arch} analyzes the PMU communication network model and provides an efficient solution based on fog computing. Then Section \ref{sec:anomaly} illustrates two anomaly detection methods fit for edge, including a detailed report of simulating experiments and performance comparison of the two selected methods in subsection \ref{sec:exp}. After the demonstration of the anomaly detection at edge, in Section \ref{sec:improve} a simulation of PMU communication network based Reiverbed Modeler will be built to evaluate the benefit in terms of network efficiency brought by the proposed PMU fog. Finally, Section \ref{sec:conclusions} concludes the paper.  

\section{Background and Related Work}
\label{sec:rel}
\subsection{Synchrophasor System and Application Requirements}
\begin{figure}
    \centering
    \includegraphics[width=0.4\textwidth]{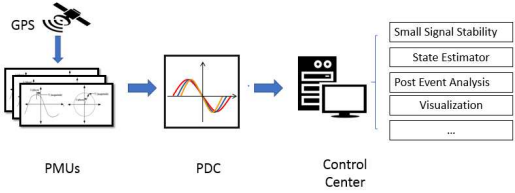}
    \caption{Synchrophasor system.}
    \label{fig:fig1}
    \vspace{-10 pt}
\end{figure}

As shown by Fig. \ref{fig:fig1}, a synchrophasor system consists of four types of basic components, the data acquisition device PMU, the phasor data concentrator (PDC), the server to operating WAMC applications and the communication networks~\cite{delivery2013synchrophasor}.  

PMUs measures the instantaneous voltage and current at specific location with the sampling rate of 2400 Hz or higher, and then convert them into phasors of 30 Hz or more. Each PMU measurement is time-stamped by the global positioning system (GPS) clock as such a PMU measurement is also called as synchrophasor. The synchrophasors measured by PMU are sent to the remote peer PDC by dedicated communications. The PDC aggregate and time-align synchrophasor data from multiple sources to create a system-wide visibility to the servers for processing in applications software.

\begin{table*}[t]
 	\caption{Application Data Requirement for WAMC}
 	\begin{center}
 	\begin{tabular}{p{2.5cm} p{5.5cm} p{1.5cm} p{2cm} p{1.5cm} p{2.25cm}}
 			\hline
 			Class& Applications based on it  & Latency & Sampling Rate &  Time Window & Accuracy\\
 			\hline
 			State Estimation  &  &1-2 Minutes & 1 samples/sec. & 5-10 Minutes &  Somewhat Important\\
 			Transient Stability & Load Trip, Generation Trip, Islanding & 100 ms & 30 samples/sec. & Few Minutes &Critically Important\\
 			Small Signal Stability  & Modes, Modes shape, Damping, Online update of PSS, Decreasing tie-line flows  &1-5 Seconds &10 samples/sec. & 10-15 Minutes & Critically Important \\
 			Voltage Stability &   &Few Seconds &30 samples/sec. & 1 hour & Critically Important \\
 			Post-Event Analysis  & Disturbance Analysis Compliance, Frequency Response Analysis, Model Validation & NA & 30 samples/sec. & Few minutes & Critically Important \\
 			Visualization & Wide-Area Monitoring and Visualization   &1-5 Seconds &1 sample/sec.& Snapshot & Not very important  \\ 			 			
 			\hline
 		\end{tabular}
 		\label{tab:wamc}
 	\end{center}
 \end{table*}

The synchrophasor system has brought benefits to the power grid with a large number of actual and potential wide area  synchrophasor applications. According to the North American Electric Reliability Corporation (NERC)~\cite{patel2010real}, the use of synchrophasor data can be classified into two general categories, real-time operations applications and planning applications (off-line applications). The former category includes Wide-Area Monitoring and Visualization (WAMV), Oscillation Detection, Frequency Stability Monitoring, Voltage Stability Monitoring, Disturbance Detection and Alarming Studies, State Estimation and so on. The later category includes Post-Event Analysis, Model Validation and so on. The North American SynchroPhasor Initiative (NASPI) provides detailed applications requirements and classification, as shown in Table \ref{tab:wamc}. There are several applications that have a strict requirement on the latency while they are not tolerant to the loss of data accuracy. For example, the application of transient stability analysis requires the latency less than 100ms. And the total latency includes the sensor processing time, the data end-to-end (ETE) delay in network, the computing time for decision making, the transmission time for control signal, and the operating time of local device. Therefore, there is a big challenge to keep a low data ETE delay in the network.

\subsection{Fog Computing}
As more and more smart mobile devices are ubiquitously deployed, it is natural to utilize the abundant computing power at the edge of network. Compared with cloud computing, the key advantage of fog computing is the on-site processing that eliminates the communication time~\cite{stojmenovic2014fog}. The attractive features such as low latency, geo-distribution and real-time interaction make fog computing an ideal platform to deliver instant services in connected sites. A three-level architecture was presented for healthcare infrastructure in which the second level is a Fog-Computing-informed-paradigm~\cite{stantchev2015smart}. Another suggested application of fog computing is augmented reality and real-time video analysis~\cite{yi2015survey}. Augmented reality applications overlay information on the real world where customers are sensitive to the latency even at the level of tens of milliseconds. In traffic surveillance and target tracking applications, cameras and sensors are intensively deployed in urban areas. Fog computing can efficiently utilize the computing and storage resource to meet the tight latency requirements~\cite{chen2017enabling}.

A number of research efforts have been reported in literature where the challenges and research directions are identified~\cite{chiang2016fog},~\cite{liu2017framework}. As to low-latency processing and resource allocation efficiency, fog computing facilitates mission critical applications at the network edge. A smart urban traffic surveillance system has been implemented with fog computing~\cite{chen2017enabling},~\cite{chen2016dynamic}. It shows that big urban data can be processed in real-time, which is essentially significant in decision making. Fog-to-cloud architectures for smart cities are explored and benefits are evaluated~\cite{ramirez2017evaluating}. The utilization of communication and computing resources enables efficient real-time services.  

Researchers also explored the potential of fog computing in transmission, storage and processing of big data in a real-time manner in a smart grid monitoring system. A data aggregation approach has been applied on devices, where fog computing enabled routers to improve the throughput for power line communication (PLC) for Smart Meters~\cite{nazmudeen2016improved}. Vehicular Delay-Tolerant Networks (VDTN) has been used to prevent data dissemination of various devices by the mobile devices such as vehicles located at the edge of the network~\cite{kumar2016vehicular}. Both of the two approaches depend on the distributed storage capability of the edge devices. 

A decentralized voltage stability monitoring method was proposed that is based on a fault-tolerant distributed computing architecture (DCBlocks)~\cite{lee2016decentralized}. A state estimation scheme was suggested that tried to take advantages of both cloud and fog computing paradigm~\cite{meloni2017cloud}. A distributed state estimation method was presented that is placed with 5G mobile cellular networks~\cite{cosovic20175g}. The three approaches addressed a particular WAMC application at the edge of the network and reduced the communication burden compared with centralized application. However, it is risky to make a decision without global information. Meanwhile, as the edge devices possess limited computing and storage resources, focusing on one specific application may sacrifice the performance on other applications. For example, it may be a trade-off to suffer longer sensor processing time.

A recently proposed hierarchical anomaly detection architecture shares the same insight with us, however, it did not tackle the delay under the umbrella of communication network, neither was the advanced edge-fog-cloud computing paradigm leveraged \cite{jamei2017anomaly}. In contrast, the proposed PMU fog focuses on prioritizing the data such that the existing QoS in communication networks can be leveraged to enable higher QoS for delay sensitive, mission critical applications. The most attractive features of the PMU fog are in two folds: (1) low cost of computational and storage resource of the edge device; and (2) no modification is required on the data packages.       

\section{PMU Fog Architecture for Efficient WAMC}
\label{sec:arch}
\subsection{PMU Fog based PMU Network}
\begin{figure}[t]
    \centering
    \includegraphics[width=0.48\textwidth]{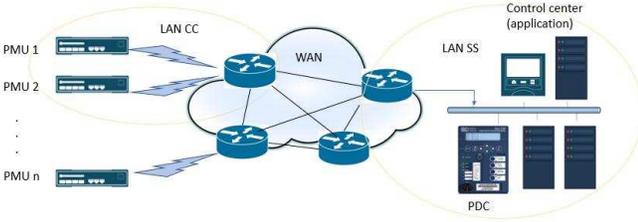}
    \caption{Synchrophasor system communication network.}
    \label{fig:fig2}
    \vspace{-10 pt}
\end{figure}

Figure \ref{fig:fig2} presents a typical communication network of the synchrophasor system \cite{chenine2011modeling}. PMUs placed in specific substations are connected to the router of the substation via a local area network (LAN), represented by \emph{LAN CC}. There may exist multiple tiers of PDCs considering multiple factors, such as the long distance, data volume, and administration issues. In this model, a simplified architecture is applied, in which it assumes a single layered PDC residing at the control center. Substation routers are able to communicate with the control center router through a wide area network (WAN). In the domain of control center, data transmission depends on another LAN, represented by \emph{LAN SS}, where the PMU measurement data are transmitted to PDC for sorting and synchronization first, and then sent upstream to multiple applications and operations.

\begin{figure}[t]
    \centering
    \includegraphics[width=0.4\textwidth]{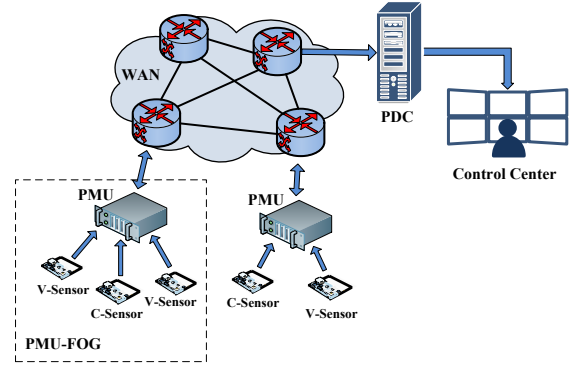}
    \caption{PMU Fog Architecture.}
    \label{fig:arch}
    \vspace{-10 pt}
\end{figure}

Figure \ref{fig:arch} shows the WAMC network architecture with the PMU fog proposed in this paper. To reduce the ETE delay without sacrificing data completeness, the PMU fog conducts anomaly detection function at the network edge. Once an anomalous activity is detected, the data packets are assigned a higher priority and then sent into the communication network. 

According to the WAMC application requirements shown in Table \ref{tab:wamc}, the latency sensitive applications are all transient stability related. It is reasonable to assume that the sensed data related to grid failures are more valuable and should be received by the controller timely. Therefore, the proposed PMU fog assigns the data packets related to grid anomalies a higher priority of transmission. To achieve this goal, the PMUs are required to detect the anomalies on site. Once an abnormal situation is detected by a PMU, it marks the current data with a higher QoS level immediately. As long as the network supports the QoS protocols, the marked data will be transmitted with higher priority. Then the PDC at control center receives the data of interest and starts processing with less time delay.

Leveraging the high priority transmission, the data reflecting the abnormal situation of the grid will receive a guaranteed bandwidth and stricter limit on allowed jitters. Consequently, a shorter ETE transmission delay and less fluctuation in latency are achieved, which means a low average $T_{WAN}$ and less opportunity of extremely long delay. Therefore, the timeout $T_{TO}$ can be set to a relatively small value with lower data loss rate. The proposed PMU fog will benefit the power grid by addressing two major factors, the ETE delay and data completeness, which are discussed in the following subsections.

\subsection{End to End (ETE) Delay}
The ETE delay counts the time from when a specific packet is created to when it arrives at the destination. The delays incurred by the transducer, phasor generation and abnormal detection will be discussed in the latter section. This section is focused on the analysis of the delay afterward. 

For a specific PMU $i, (i=1,2,...,N)$, its ETE transmission delay $T_i$ is,
\begin{equation}
T_i=T_{CCi}+T_{WANi}+2T_{SSi}+T_{PDC}
\end{equation}

\noindent{where $T_{CCi}$, $T_{WANi}$ and $T_{SSi}$ are the transmission time on \emph{LAN CC}, \emph{WAN}, and \emph{LAN SS}, and $T_{PDC}$ is the data processing time on PDC. It is reasonable to assume that $T_{CCi}\simeq T_{SSi}<<T_{WANi}$, then:}

\begin{equation}
T_i=T_{WANi}+T_{PDC}
\end{equation}

Therefore the ETE delay largely depends on the transmission time on the \emph{WAN} and the speed of PDC. $T_{WANi}$ is determined by the physical distance and network condition. More analysis of $T_{PDC}$ is as the following.

Besides functions like data quality check and data storage, the basic function of PDC is data alignment with time-stamps, data synchronization. To align data according to their time stamps, the PDC will gather the PMU measurements with the same time stamp into one set. Each PDC assigns a timeout $T_{TO}$ for each time stamp and this timeout would be the amount of time that the buffer waiting for the PMU measurements with this particular time stamp arriving at PDC. When the set is full or the timeout expires, the PDC will forward the data in the buffer to applications and be ready for the data with next time stamp. The timeout $T_{TO}$ is typically user configurable and set according to the expected worst-case network communications delay. Hence, the PDC processing time $T_{PDC}$ is determined by the timeout $T_{TO}$ and the transmission time in \emph{WAN} of each PMU $T_{WANi}$:

\begin{equation}
T_\theta=max(T_{WANi}) \qquad i= 1,2,...,N
\end{equation}
\begin{equation}
T_{PDCi}=\{
\begin{array}{c}
T_\theta-T_{WANi}, \qquad\mathrm{if}\ T_\theta<T_{TO}, \\
T_{TO}-T_{WANi}, \qquad\mathrm{if}\ T_\theta\geq T_{TO}.\\
\end{array}
\end{equation}

\subsection{Data Incompleteness}
Data incompleteness is another critical issue in PMU communication networks. It directly influences the data quality in upper layer applications and further application accuracy. A timeout $T_{TO}$ is assigned to make sure that maximum amount of synchronized data are received. However, when a packet is lost or an extremely long transmission delay occurs, the corresponding measurement cannot arrive PDC in time, and the data incompleteness is inevitable during the data synchronization. Regardless of the packet loss, increasing the $T_{TO}$ can effectively reduce the data incompleteness resulted from long transmission delay. However, longer $T_{TO}$ directly lead to a longer ETE delay. Obviously, it is a trade-off between the shorter ETE delay and the lower data incompleteness rate.

\section{Anomaly Detection by PMU Fog}
\label{sec:anomaly}
As discussed above, the capability of conducting anomaly detection tasks at the edge is the key. In this work two anomaly detection approaches, Singular Spectrum Analysis (SSA)~\cite{polunchenko2012state} and K-nearest neighbors (KNN)~\cite{jain1999data}, are investigated for the feasibility to be implemented on PMU fog. It is challenging to migrate computing intensive tasks to the fog due to constraints at the network edge, such as limited computing resource and lack of global information. 

\subsection{SSA based Anomaly Detection}
SSA is a model-free and structure-less approach. It separates the sampled time series into multiple components, and then analyzes a linear recurrence relation (LRR) for interpretable components. By selecting different categories of components as research objects, SSA has been recognized in a wide field of time series processing applications, such as finding data structure, extracting periodic pattern and complex trends, smoothing and change point detection.

The SSA based anomaly detection is actually a quick change point detection method. An important assumption of this method is that, a fault or event always starts from a normal condition. When a fault or an event occurs, the measurement data will experience a sudden change from normal to abnormal. The abnormal detection based on SSA is to catch this particular change.

The basic SSA algorithm can be described as the following. Assume $\mathbb{X}=(x_1,x_2,\cdots,x_N)$ is a period of PMU measurement series with the length of $N$. The original time series $\mathbb{X}$ is truncated by a sliding window, with a fixed window length of $M$, into a series of lagged vectors $\overrightarrow{X_i}$, as in Eq.~\ref{eq:vector}. These vectors are transformed to a trajectory matrix $X$, in size of $M\times K$, $K=N-M+1$, as in Eq.~\ref{eq:traj}. The trajectory includes the whole information of the original time series $\mathbb{X}$.

\begin{equation}
  \overrightarrow{X_i}={\left(x_i,\cdots,x_{M+i-1}\right)}' , i=1,\cdots,K
  \label{eq:vector}
\end{equation}
\begin{equation}
X=\left[\overrightarrow{X_1},\overrightarrow{X_2},\cdots,\overrightarrow{X_K}\right]=\left(x_{i,j}\right)_{i,j=1}^{M,K}
\label{eq:traj}
\end{equation}

Columns $X_j$ of the trajectory matrix $X$ can be considered as vectors in an $M$-dimensional space $\mathbf{R}_M$. A particular combination of a certain number $l$ of the Singular Value Decomposition (SVD) eigenvectors determines an $l$-dimensional subspace $\mathcal{L}_l$ in $\mathbf{R}_M$, $l<M$. The M-dimensional data ${X_1,\cdots,X_K}$ is then projected onto the subspace $\mathcal{L}_l$.

Apply SVD procedure on the trajectory matrix $X$ (suppose the rank of $X$ is $d$) and obtain $d$ collection sets $(\sqrt{\lambda_i}, U_i,V_i)$, $i=1,2,\cdots,d$, which is called $i$-th eigentriple of the SVD. The eigentriples include, $d$ non-zero singular values $\sqrt{\lambda_1},\sqrt{\lambda_2},\cdots,\sqrt{\lambda_d}$ in decreasing order and the corresponding left singular vectors $U_1,U_2,\cdots,U_d$, and right singular vectors $V_1,V_2,\cdots,V_d$.  Each eigentriple represents one direction of projection, and each eigentriple $(\sqrt{\lambda_i}, U_i,V_i)$ can rebuild a rank-one matrix $X_i=\sqrt{\lambda_i} U_i V_i'$. The $X_i$ are bi-orthogonal to each other, and compose the basis set of the trajectory matrix $X$. The eigenvalue $\sqrt{\lambda_i}$ represents the level of contribution of the specific basis matrix $X_i$ to the trajectory matrix, i.e., the level of information contained in matrix $X_i$.

After decomposing the trajectory matrix, the next step is selecting a subset indices $I$ of basis eigentriples $(\sqrt{\lambda_i}, U_i,V_i)$ or basis matrices $X_i$ to the $l$-dimensional subspace $\mathcal{L}_l$ in $\mathbf{R}_M$, where $l<M$.

\begin{equation}
I=\lbrace i_1,i_2,\cdots,i_l \rbrace
\end{equation}

Diagonal averaging is used to transfer matrix $X_{I}=\sum_{i\in I} X_i $ into a time series $\mathbb{X}_I$ (reconstruction), which is the component-sum of the original series $\mathbb{X}$.

\begin{equation}
\mathbb{X}_I(i)=
\left\{
\begin{array}{c}
\frac{1}{i}\sum_{j=1}^{i}x_{j,i-j+1}\qquad \qquad \mathrm{for}\ 1\le i<M\\
\frac{1}{M}\sum_{j=1}^{M}x_{j,i-j+1}\qquad \qquad \mathrm{for}\ M\le i\le K\\
\frac{1}{N-i+1}\sum_{j=i-K+1}^{N-K+1}x_{j,i-j+1}\ \mathrm{for}\ K< i\le N\\
\end{array}
\right.
\label{Eq:model}
\end{equation}

With deliberate selection of $I$ (usually the first $l$ eigenvalues), the reconstruction time series will be reasonably close to original series $\mathbb{X}$. The change-point detection is conducted in the following steps. Specify one section of target time series, use SSA to compute its $l$-dimensional reconstructed matrix $X_{I}$, and observe the distance between the l-dimensional matrix $X_{I}$ and an embedded matrix of the test time series. Then the change point is supposed to be the time point when a significant increasing of distance is observed.

\vspace{1mm}
\noindent{\textbf{Step 1: Construction of target matrix}}

Firstly, embed the target time series $\overrightarrow{x^{(n)}_{target}}$ into the target matrix $X^{(n)}_{target}$,

\begin{equation}
\overrightarrow{x^{(n)}_{target}}=(x_{n+1},x_{n+2},...,x_{n+N})
\end{equation}
\begin{equation}
X^{(n)}_{target}=[\overrightarrow{X_1},\overrightarrow{X_2},...,\overrightarrow{X_n}]
\end{equation}
\begin{equation}
\overrightarrow{X_i}={\left(x_{n+i},\cdots,x_{M+n+i-1}\right)}',\ i=1,\cdots,K 
\end{equation}
where $K=N-M+1$. Note that $X$ is a Hankel matrix, which has the equal elements on it's skew-diagonals $i+j=const$. Apply SVD to target matrix $X^{(n)}_{target}$, and get all of the eigentriples $(\sqrt{\lambda_i}, U_i,V_i)$, $i=1,2,\cdots,d$. Then, select the first $l$ eigentriples, and build the $l$-dimensional reconstructive target matrix $X^{(n)}_{target,l}$.

\vspace{1mm}
\noindent{\textbf{Step 2: Construction of test matrix}}

\begin{equation}
\overrightarrow{x^{(n)}_{test}}=(x_{n+1+p},x_{n+2+p},...,x_{n+M+q+1})
\end{equation}
\begin{equation}
X^{(n)}_{test}=[\overrightarrow{X_1},\overrightarrow{X_2},...,\overrightarrow{X_{q-p-1}}]
\end{equation}
\begin{equation}
\overrightarrow{X_i}={\left(x_{n+p+i},\cdots,x_{M+n+p+i-1}\right)}',\ i=1,\cdots,q-p+1. 
\end{equation}

So the size of the test matrix is $M\times (q-p+1)$.

\vspace{1mm}
\noindent{\textbf{Step 3: Computing distance}}

The difference between above two matrices is measured by the sum of the squared Euclidean distance.
\begin{equation}
\mathcal{D}_{n,I,M,p,q}=\sum\limits_{i=p+1}^{M+q+1}{((\overrightarrow{X^{(n)}_{test,i}})'\overrightarrow{X^{(n)}_{test,i}}-(\overrightarrow{X^{(n)}_{test,i}})'UU'\overrightarrow{X^{(n)}_{test,i}})}
\end{equation}
Note that the distance calculated in this equation describes the condition at time point $t=n+M+q+1$.

\subsection{KNN based Anomaly Detection}
The idea of KNN based abnormal detection is pretty straightforward. Basically, the input is classified by a majority vote of the $k$ nearest training examples in the feature space. First, all training data with classification labels are mapped into the $l$-dimension feature space, where $l$ is the number of selected features. To classify a specific data point, the Euclidean distance in feature space between the test data point with each training data is calculated. Then the $k$ nearest neighbors are identified as well as the category of the majority. In the end, the test data point is assigned to the category of the majority of its $k$ nearest neighbors. 


The feature extraction methodology processes the sampled voltage and/or current signals using windows of a size that is equivalent to one cycle acquired through PMU device. In total 16 features are obtained in both time domain and frequency domain. The time domain features can be divided into three groups. The first group summarizes the window in a global view of the data: harmonic mean ($F_1$), standard deviation ($F_2$), mean deviation ($F_3$), and Kurtosis ($F_4$).

\begin{equation}
F_1=\frac{N}{\sum\limits_{j=1}^{N}\frac{1}{d_j}}
\end{equation}
\begin{equation}
F_2=\sqrt{\frac{\sum\limits_{j=1}^{N}(d_j-mean)^2}{N-1}}
\end{equation}
\begin{equation}
F_3=\frac{\sum\limits_{j=1}^{N}|d_j-mean|}{N}
\end{equation}
\begin{equation}
F_4=\frac{\frac{1}{N}\sum\limits_{j=1}^{N}(d_j-mean)^4}{[\frac{1}{N}\sum\limits_{j=1}^{N}(d_j-mean)^2]^2}
\end{equation}

The second group of features performs entropy calculations for the aforementioned window. The entropy is originally defined to work with random variables, but it describes the confusion degree of data in the aforementioned window. The features in this category are entropy ($F_5$), Shanon entropy ($F_6$) and Renyi entropy ($F_7$). $\alpha$ is a variable used in calculating Renyi entropy, which is adjusted by a variety tests and the best value is 0.4.

\begin{equation}
F_5=\sum\limits_{j=1}^{N}log(d_j^2)
\end{equation}
\begin{equation}
F_6=-\sum\limits_{j=1}^{N}d_j^2\times log(d_j^2)
\end{equation} 
\begin{equation}
F_7=\frac{1}{1-\alpha}\times log\sum\limits_{j=1}^{N}d_j^\alpha
\end{equation}

The third group of features describes amplitude of the signal. This group includes root mean square value (RMS, $F_8$), peak value ($F_9$) and the difference between the maximum and the minimum amplitudes of the signal ($F_{10}$).

\begin{equation}
F_8=\sqrt{\frac{1}{N}\sum\limits_{j=1}^{N}|d_j|}
\end{equation}
\begin{equation}
F_9=max(d_j)
\end{equation}
\begin{equation}
F_{10}=max(d_j)-min(d_j)
\end{equation}

The rest of the features are obtained in the frequency domain. The unique features in frequency domain are the harmonic components and obtained by fast Fourier transform (FFT). Note that FFT is already implemented in the hardware in PMU to generate phasors, so the computation cost is low. The FFT is calculated as:
\begin{equation}
D_k=\sum\limits_{j=1}^{N}{d_je^{-j2\pi k\frac{j}{N}}}
\end{equation}

The feature $F_{11}$ is the total harmonic distortion (THD) of the frequency components $D_k$:
\begin{equation}
F_{11}=\frac{\sqrt{\sum\limits_{k=2}^{N}D^2_k}}{D_1}
\end{equation}

The features $F_{12}$ to $F_{16}$ are the amplitudes of the fundamental and harmonic components of second to fifth orders.


After feature extraction, a classifier determines whether or not the operation condition of the current period is normal. Considering the computing complexity at PMU, KNN is a good fit due to its simplicity.

\subsection{Experimental Study and Performance Evaluation}
\label{sec:exp}
In order to evaluate and compare the detection performance and the computational cost of the SSA based and the KNN based anomaly detection methods, an IEEE 16-machine 68-bus system~\cite{graham2000power} is built using Power System Toolbox (PST) for a numerical simulation. Five representative fault types are employed, which are generator trip (GT), line-to-line fault (LL), line to ground fault (LG), line-to-line to ground fault (LLG), and three phase fault. The simulation of each fault record lasts 2s with time solution of 0.0017 ($1/600$)s. Each fault randomly occurs at 0.3, 0.4, 0.5, 0.6, and 0.7s with duration for 0.01s (6 cycles). The location of each fault is also randomly distributed over the whole system area. The PMU's location is fixed at one of the 68 buses. The voltage measurement record are used for detection.

\subsubsection{Detection Accuracy}
First of all, the detection accuracy on different fault categories is studied. For each category, 37 records are simulated respectively. The ambient (load) noise level is set to 5\%.   

For SSA, the selection of parameters is $N=36, M=18, p=18, q=30$ and $l=6$. To keep computing cost low, the sampling rate is reduced by five times, i.e., the time resolution of dataset for SSA is 0.0083 ($1/120$)s. The distance calculated for each time point is normalized and the threshold is selected as the $9\times 10^6$ times of the average distance when the measurement signals describe a normal condition.

For KNN, the number of the interested neighbors is five. The window length is ten, and the time resolution is 0.0017s and the window length is equivalent to one cycle. Since KNN classification algorithm is an instance based learning algorithm, five fault records for each type of fault, i.e., 595 instances in feature space in total are randomly selected for training. And the rest of 32 fault records are utilized as testing data. 
Four features, Shanon entropy ($F_6$), RMS ($F_8$), peak value ($F_9$) and the difference between the maximum and the minimum amplitudes of the signal ($F_{10}$), are selected as inputs to pursue a balance of computational cost and accuracy.  

\begin{figure}[t]
    \centering
    \includegraphics[width=0.35\textwidth]{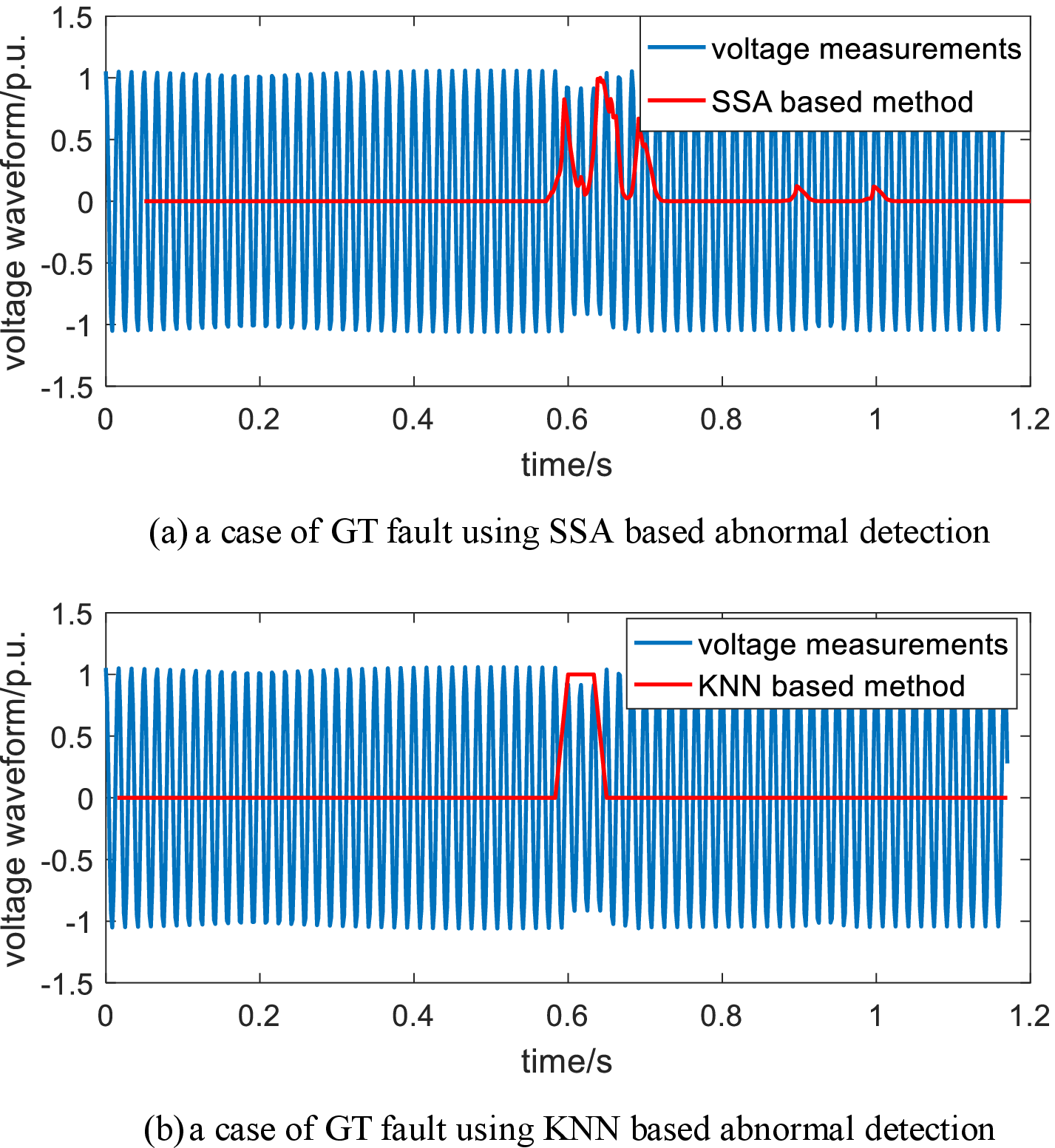}
    \caption{A case of GT fault.}
    \label{fig:gt}
    \vspace{-10 pt}
\end{figure}

\begin{figure}
    \centering
    \includegraphics[width=0.35\textwidth]{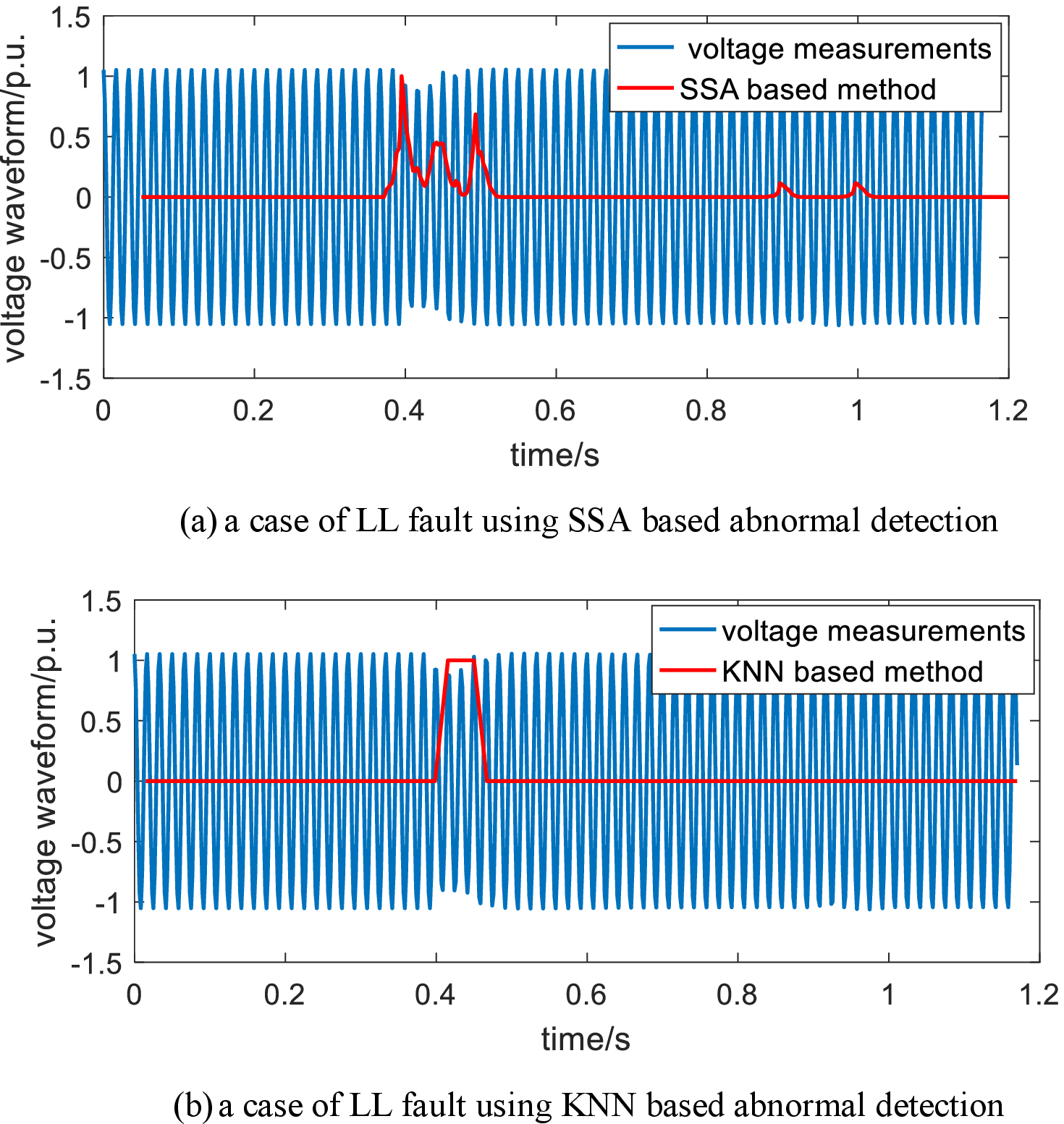}
    \caption{A case of LL fault.}
    \label{fig:ll}
    \vspace{-10 pt}
\end{figure}

\begin{figure}
    \centering
    \includegraphics[width=0.35\textwidth]{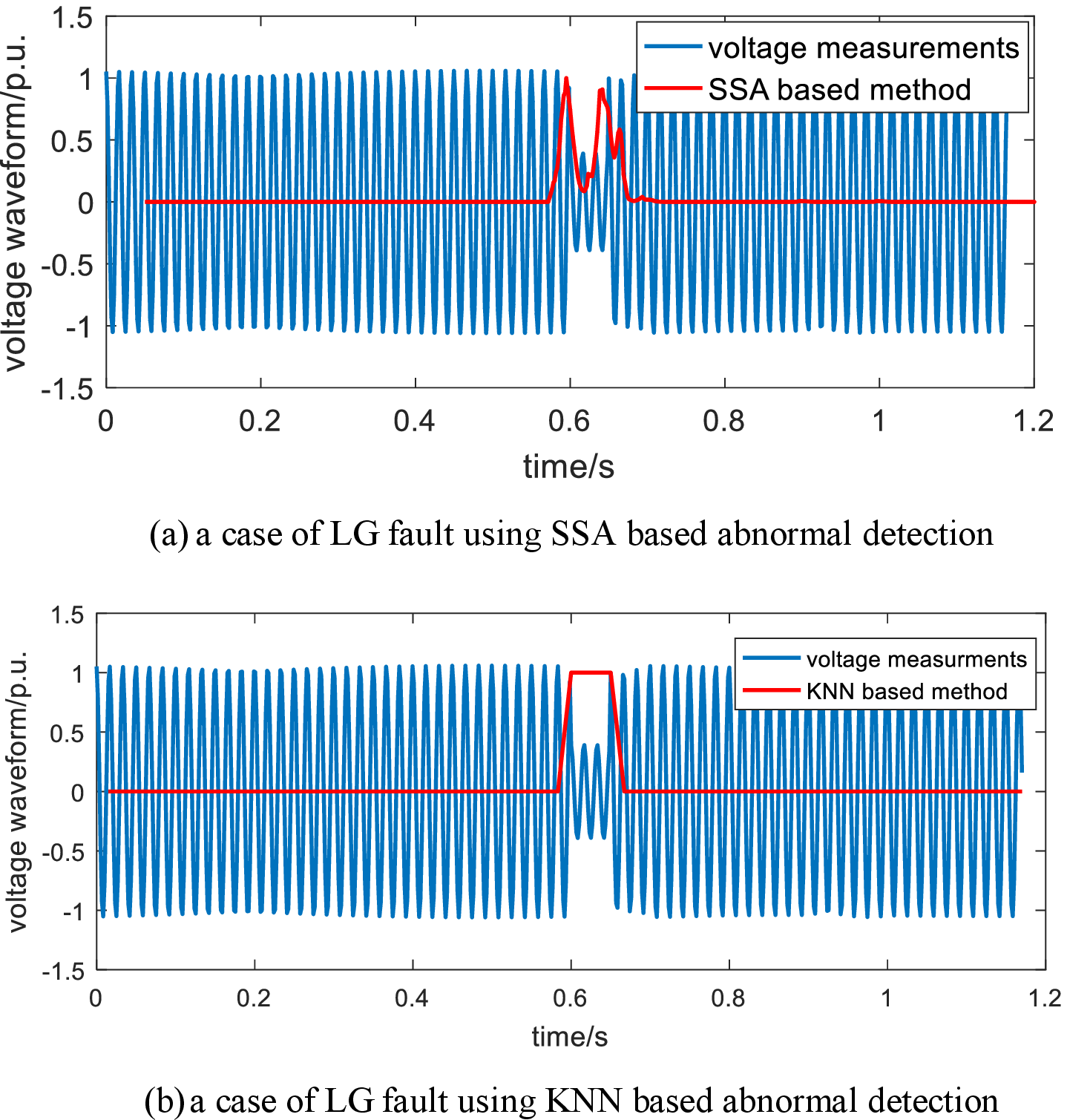}
    \caption{A case of LG fault.}
    \label{fig:lg}
    \vspace{-10 pt}
\end{figure}

\begin{figure}
    \centering
    \includegraphics[width=0.35\textwidth]{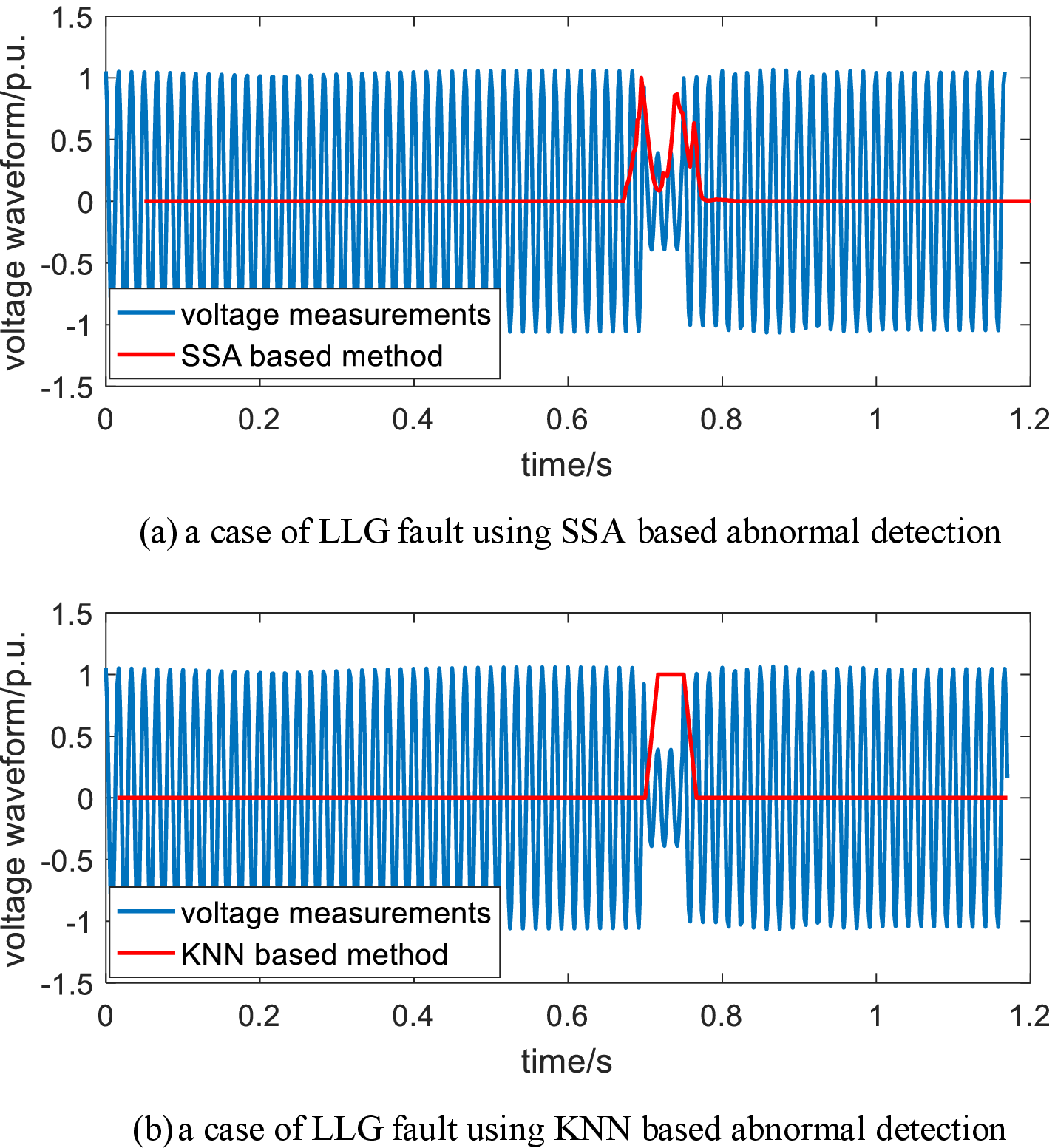}
    \caption{A case of LLG fault.}
    \label{fig:llg}
    \vspace{-10 pt}
\end{figure}

\begin{figure}
    \centering
    \includegraphics[width=0.35\textwidth]{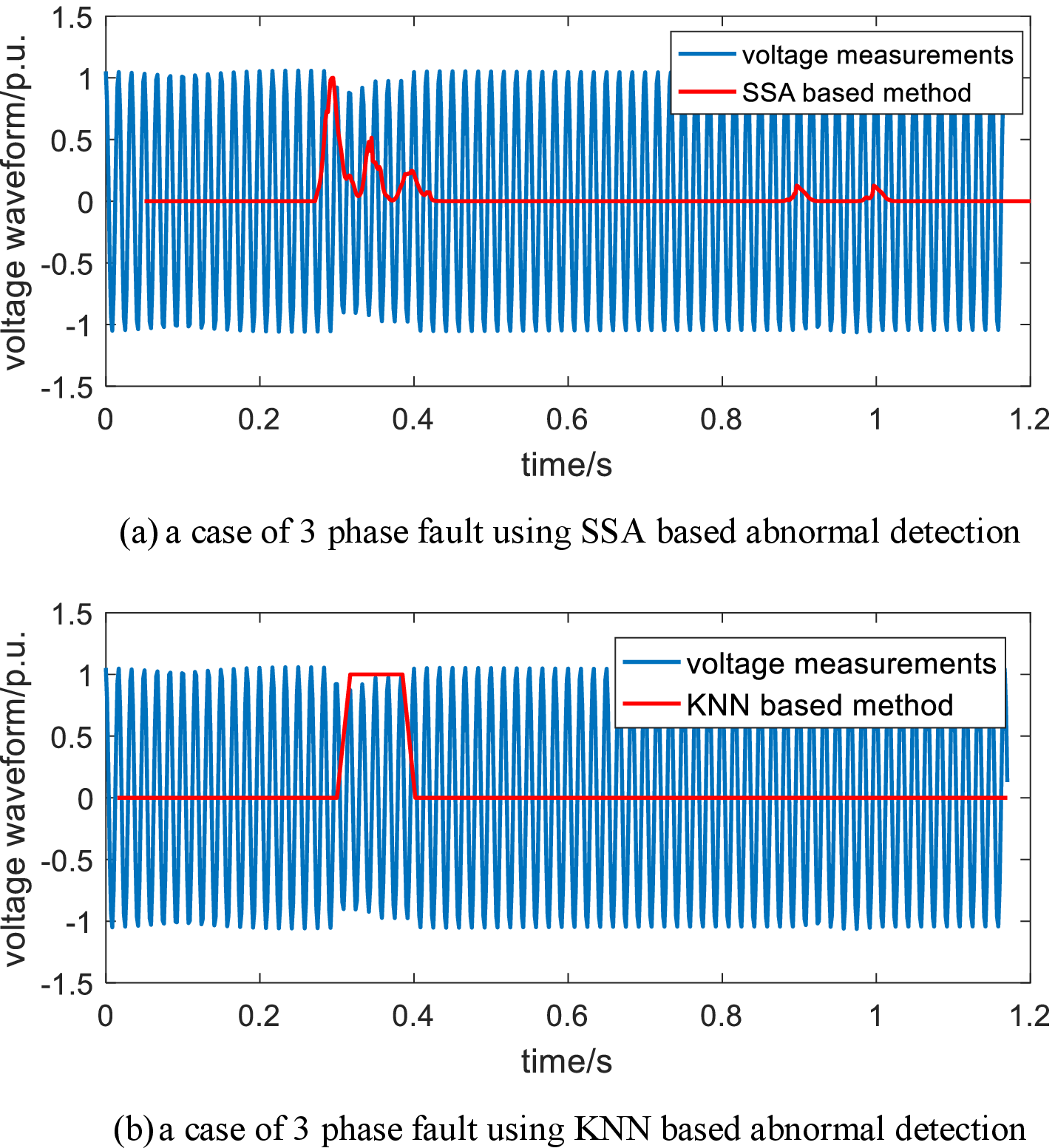}
    \caption{A case of three-phase fault.}
    \label{fig:3p}
    \vspace{-10 pt}
\end{figure}

Figures \ref{fig:gt}, \ref{fig:ll}, \ref{fig:lg}, \ref{fig:llg}, \ref{fig:3p} provide examples of anomaly detection for each type of faults respectively, where we can see the detail of detection result. Both the SSA and the KNN successfully identified the fault. While the SSA method is more sensitive to the marginal fluctuation of the voltage measurement, the detection of the KNN method always has a delay about one cycle, which is incurred by the window length. 

Tables~\ref{tab:ssa} and~\ref{tab:knn} present the performance on different fault categories by the SSA method and KNN method respectively. The SSA method, although has higher false positive rate (FPR) due to its sensitivity to small changes, generally outperforms the KNN method.

\begin{table}[htb!]
	\caption{Detection accuracy of the SSA method with 5\% noise}
	\begin{center}
		\begin{tabular}{ccc}
			\hline
			&True positive rate  & False positive rate  \\
			\hline
			LL  &100\%  &0\\
			LG  &100\%  & 0  \\
			LLG &100\%  &3.13\% \\
			3Phase  &100\% &6.25\% \\
			GT  &100\%  &18.75\% \\
			\hline
		\end{tabular}
		\label{tab:ssa}
	\end{center}
	\end{table}

\begin{table}[htb!]
	\caption{Detection accuracy of the KNN method with 5\% noise}
	\begin{center}
		\begin{tabular}{ccc}
			\hline
			&True positive rate  & False positive rate \\
			\hline
			LL  &90.63\%  &3.13\%\\
			LG  &90.63\%  & 0  \\
			LLG &93.75\%  &0 \\
			3Phase  &87.50\% &0 \\
			GT  &78.50\%  &3.13\%\\
			\hline
		\end{tabular}
		\label{tab:knn}
	\end{center}
\end{table}

\subsubsection{Robustness on Different Noise Levels}
The performances of the SSA method and KNN method are also compared against noise levels. The testing ambient noise levels are set as 3\%, 5\% and 7\%, which are close to the real situation in power systems. For each noise level, 140 fault records including all five fault categories are randomly generated. The simulation of each fault record lasts 2 seconds and each fault randomly occurs at 0.3, 0.4, 0.5, 0.6, and 0.7 seconds with duration for 0.01s (6 cycles).  

\begin{figure}
    \centering
    \includegraphics[width=0.35\textwidth]{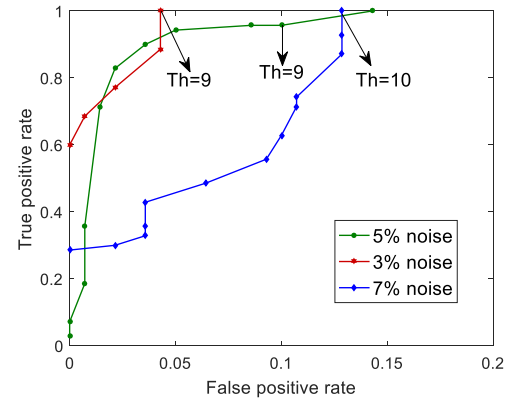}
    \caption{Detection performance of SSA on different noise levels.}
    \label{fig:roc_ssa}
    \vspace{-10 pt}
\end{figure}

Figure \ref{fig:roc_ssa} shows the Receiver Operating Characteristic (ROC) curves of SSA with various threshold settings. It is not a surprise that the lower the noise level is, the better the performance is. Consider each curve of different noise levels individually, the optimal detection result with noise levels of 3\%, 5\% and 7\% are with TPR \& FPR of 100\% \& 4.29\%, 94.29\% \& 5.00\% and 100\% \& 12.86\%, respectively, as marked in Fig. \ref{fig:roc_ssa}.

\begin{figure}
    \centering
    \includegraphics[width=0.35\textwidth]{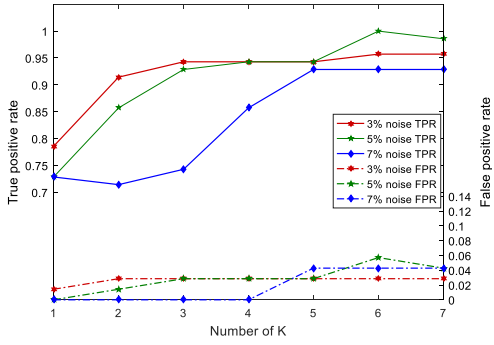}
    \caption{Detection performance of KNN on different noise levels.}
    \label{fig:roc_knn1}
    \vspace{-10 pt}
\end{figure}

KNN is an instance-based learning method, so it is not possible to plot a ROC curve with continuously threshold changing. Alternatively, TPR and FPR with different number of training data sets are used. Figure~\ref{fig:roc_knn1} shows the detection result with training data set number from one to seven, where the training data sets are delicately selected. Since the training data set size directly determines the method complexity, the trade-off of the detection performance and the computational cost is also investigated. With five selected data sets, the TPR \& FPR for noise level 3\%, 5\% and 7\% are 94.29\% \& 2.86\%, 94.29\% \& 2.86\% and 92.86\% \& 4.29\%. 

\begin{figure}
    \centering
    \includegraphics[width=0.35\textwidth]{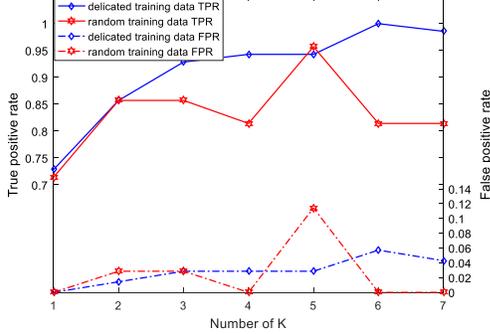}
    \caption{Comparison of different training data sets.}
    \label{fig:roc_knn}
    \vspace{-10 pt}
\end{figure}

As KNN is instance-based, the performance largely depends on the quality of training data. Figure~\ref{fig:roc_knn} gives an example of an random training data set comparing with the selected one with noise level 5\%. And it is obvious that the performance of KNN is not stable with randomly selected training data set.   

\subsubsection{Computational Complexity Analysis}
To verify the feasibility of our proposed PMU fog scheme in wide area smart grid anomaly detection, the algorithms are implemented and tested on two different computing devices. 
One is a laptop with an Intel i7-7700HQ processor, 2.80 GHz processing frequency and 8 GB RAM memory. The averaged running time of KNN approach is 0.0123s and the averaged running time of SSA algorithm is 0.0152s. Another devices is Raspberry Pi 3 Model B. Its processor is 64/32-bit quad-core ARM Cortex-A53 and it has a processing frequency of 1.2 GHz. The memory is 1 GB. The KNN algorithm runs an averaged time 0.1512s on it while the averaged running time of SSA is 0.1896s.

\subsubsection{Discussions} 
In general, both of the two anomaly detection methods can serve the purpose at the PMU device and are able to detect anomalies accurately. With their own advantages and constraints, they can be applied alternatively in practice according to actual working environments. More specifically, the advantages of the SSA method include higher detection accuracy, shorter detection delay, robust to different noise levels and different fault types and no need of historian data for training. In contrast, the KNN method is a better choice when low computing cost is a dominant factor, or additional information of feature extraction for other applications.

\section{Improvement in Network Efficiency}
\label{sec:improve}
A simulation model is built based on Riverbed Modeler~\cite{modeler2014riverbed} to evaluate the effects of the proposed PMU fog to the entire PMU communication network efficiency. Riverbed Modeler (used to name as OPNET) is a robust software specialized at network and system simulation, which provides a large amount of models of standard communication devices and protocols. The version of software we used in this paper is Riverbed Modeler Academic Edition 17.5.

\subsection{PMU communication Model in Riverbed}

The communication model implemented on Riverbed is based a hypothetic locations of seven PMUs in the northeast of US. Note that, this synchronphasor system model including seven PMUs does not represent the actual synchrophasor system working in northeast, but is a model in equivalent geographic distance level and network architecture. Since in a wide area network, the distance between network devices is a crucial factor for the distance-based delay, the placements of the network components are important.

\begin{figure}
    \centering
    \includegraphics[width=0.35\textwidth]{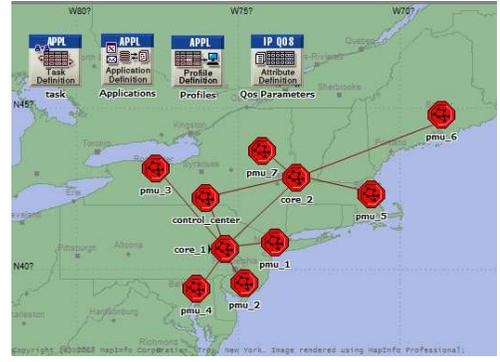}
    \caption{PMU communication network on Riverbed.}
    \label{fig:pmu_opnet}
    \vspace{-10 pt}
\end{figure}

Figure \ref{fig:pmu_opnet} is a global view of the geographical topology of the communication network model. The octagons on the topological graph stand for subnets. The subnets \textit{PMU\_i} represent LANs at substations where PMU devices implemented. In each \textit{PMU\_i} subnet, PMU device is connected to  the outer network via the substation router. The PMU device is simulated by \textit{Ethernet Workstation} model and it will generate data traffic to its destination PDC. The subnets \textit{core\_1} and the subnet \textit{core\_2} represent a network of meshed routers. In the subnets \textit{control\_center}, PDC and WAMC are configured by \textit{Ethernet Server} model. They connect with each other through a switch and connect to the wide area network with a router.  

\begin{figure}
    \centering
    \includegraphics[width=0.30\textwidth]{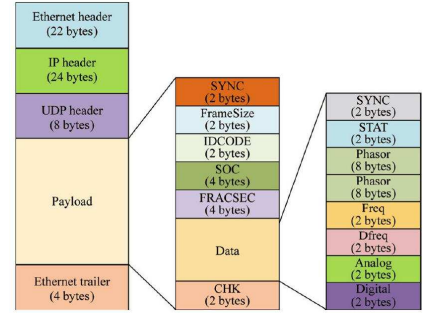}
    \caption{Data packet format for IEEE C37.118.2 protocol.}
    \label{fig:data}
    \vspace{-10 pt}
\end{figure}

The data communication requirements of synchronphasor system are well addressed by IEEE C37.118 standard \cite{martin2014overview}. The IEEE C37.118.2 defines the data packet formats for four types of messages, data, configuration, header and command. In this paper we mainly focus on the data packets, which correspond to the PMU measurements. Figure \ref{fig:data} shows the data packet structure and the total size of a data packet is 112 Bytes.  
	
The data format supports efficient and secure message exchange over the network. A paradigm of network protocols used in various layer is given in Table~\ref{tab:c37}. PMU devices in substation are constantly sending data packet to the PDC, so Constant Bit Rate (CBR) is appropriate as an application layer protocol. User Datagram Protocol (UDP) is selected as the protocol in transportation level instead of Transportation Control protocol (TCP) due to consideration of latency. Also given the latency and bandwidth requirements, the optical fibers is a popular physical media in power grid communication networks.	   
 \begin{table}[htb]
 	\caption{A paradigm for communication model(C37.118)}
 	\begin{center}
 		\begin{tabular}{cc}
 			\hline
 			\textbf{Layer}  & \textbf{Protocol} \\
 			\hline
 			Application  &CBR\\
 			Transportation  &UDP  \\
 			Network  &IP  \\
 			Data link &Ethernet  \\
 			Physical  & Optical fiber  \\	 			
 			\hline
 		\end{tabular}
 		\label{tab:c37}
 	\end{center}
 \end{table} 
 In the simulation, there is no existing application in accordance with the IEEE C37.118 standard. We used \textit{Task Config} model to customize an task generating data packets obeying the IEEE C37.118 standard. And then the \textit{Application Config} model employs this task to configure an application with the transport protocol of UDP. To mimic a PMU device, the \textit{Ethernet Workstation} will generates data message constantly with the speed of 30 packets/s and the size of 112 Bytes.

In order to transmit the phasor measurements of interest with high priority, a packet scheduling mechanism named Weighted Fair Queuing (WFQ) is used in Riverbed to support the QoS guaranteed application. Basically, WFQ assigns the data flow with high QoS requirement more bandwidth sharing than others. Therefore, application class, weight and buffer allocation are important for WFQ mechanism. In the simulation, we made WFQ classified by DiffServ. The data flow is sorted by the corresponding DiffServ Code Point (DSCP) at the data source, and assigned by specific weight and buffer size. In our case, the DSCP for all data flow is set to AF (Assured Forwarding) 23 for the scenario without QoS control, which means all of the phasor measurement data will be transmitted fairly. While in the scenario with QoS control, the phasor measurement data of interests are transmitted utilizing an application with the DSCP of EF (Expedited Forwarding), and the phasor measurement data of normal situation still use the service of AF 23. Note that EF class is the highest class with lower delay and loss compared with AF 23.    
	
In the simulation, three scenarios varying in the configuration of communication links are designed. Two of them belong to dedicated network paradigm, where the network is assigned to the utilization of PMU communication only and each PMU device has a dedicated channel connecting to the PDC. These two scenarios are different in the bandwidth assignment. \emph{Scenario 1} builds network with 10 Mbps communication links, while \emph{Scenario 2} builds network with 100 Mbps communication link. And \emph{Scenario 3} simulates the situation that PMU system shares network with traffic from other power grid devices like RTUs and IEDs. In this scenario, Ethernet 100BaseT serves as the communication link but a background traffic \textit{T3\_1hours\_bps} is added to every links from substation router to WAMC control center in the network. This background traffic constantly generates data in 45 Mbps, therefore, the background traffic level is 45\%.  
	
\subsection{Simulations Result and Discussions} 
In the simulation, the ETE transmission delay statistics are collected. The ETE transmission delay includes (1) the delay between all PMU devices and the PDC, and (2) the delay between the PDC to the server where the WAMC applications are executed. The later is much smaller value than the previous one, which is in accordance with the model analysis. For each PMU device, the ETE delay is calculated by the average value of the delay from itself to the PDC adding the average value of the the delay from the PDC to the application server, and the later one is uniform for all of the PMUs. 
Tables~\ref{tab:ete10},~\ref{tab:ete100}, and ~\ref{tab:ete100b} show the simulation result of three scenarios. Table~\ref{tab:ete10} represents the scenario of a dedicated network with 10 Mbps communication links (\emph{Scenario 1}), Table~\ref{tab:ete100} stands for the scenario of a dedicated network with 100 Mbps communication links (\emph{Scenario 2}). In Table~\ref{tab:ete100b}, 45\% background traffic is added to the 100 Mbps communication links (\emph{Scenario 3}). Generally speaking, the delay is distance-dominated. The longer distance between the PMU to the PDC, the longer ETE delay. Meanwhile, the network with 10 Mbps links suffers a higher average delay than the network with 100 Mbps because of the bandwidth limitation. While the sharing communication in the third scenario does not lose too much on delay, which may be because the bandwidth of 100 Mbps is large enough for the PMU traffic and the performance won't sacrifice too much for sharing channel. 
\begin{table}[htb]
	\caption{ETE transmission delay with 10Mbps dedicated network (ms)}
	\begin{center}
		\begin{tabular}{ccc}
			\hline
			&No QoS control  & With QoS control \\
			\hline
			PMU 1  &4.7  &3.1\\
			PMU 2  &4.8  &3.0  \\
			PMU 3  &5.1  &3.9  \\
			PMU 4  &5.5  &3.7  \\
			PMU 5  &8.9  &4.7  \\
			PMU 6  &11.6  &5.6  \\
			PMU 7  &10.5  &4.6  \\ 			 			
			\hline
		\end{tabular}
		\label{tab:ete10}
	\end{center}
\end{table}

 \begin{table}[htb]
 	\caption{ETE transmission delay with 100Mbps dedicated network (ms)}
 	\begin{center}
 		\begin{tabular}{ccc}
 			\hline
 			&No QoS control  & With QoS control \\
 			\hline
 			PMU 1  &5.4  &2.7\\
 			PMU 2  &5.2  &2.6  \\
 			PMU 3  &6.6  &5.0  \\
 			PMU 4  &3.8  &2.8  \\
 			PMU 5  &4.7  &3.4  \\
 			PMU 6  &5.8  &4.7  \\
 			PMU 7  &5.1  &3.6  \\ 			 			
 			\hline
 		\end{tabular}
 		\label{tab:ete100}
 	\end{center}
 \end{table}

\begin{table}[htb]
	\caption{ETE transmission delay with 100Mbps background communication (ms)}
	\begin{center}
		\begin{tabular}{ccc}
			\hline
			&No QoS control  & With QoS control \\
			\hline
			PMU 1  &4.7  &2.6\\
			PMU 2  &4.1  &2.8  \\
			PMU 3  &6.5  &4.9  \\
			PMU 4  &5.2  &2.8  \\
			PMU 5  &4.1  &3.1  \\
			PMU 6  &5.9  &4.7  \\
			PMU 7  &4.7  &3.6  \\ 			 			
			\hline
		\end{tabular}
		\label{tab:ete100b}
	\end{center}
\end{table}
In all scenarios, the PMU ETE transmission delays without a QoS control are all higher than that with a QoS control, especially in the bandwidth limited situation (\emph{Scenario 1}). A significant reduction of ETE transmission delay is achieved if the routers know which data packets are more important than others and apply QoS control policies. Therefore, if the electrical abnormal situation is successfully detected by PMU fog, the data of interests can be transmitted with higher priority, which means a shorter transmission delay.      
\begin{figure}
    \centering
    \includegraphics[width=0.35\textwidth]{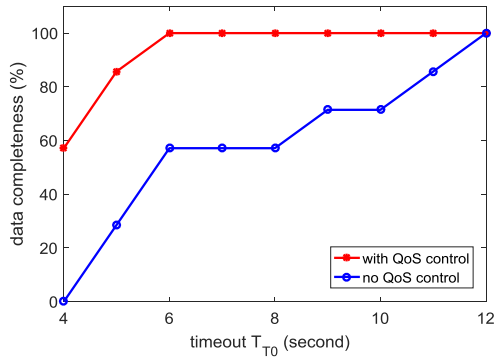}
    \caption{Data completeness with 10 Mbps dedicated network.}
    \label{fig:tt0}
    \vspace{-10 pt}
\end{figure}
The data completeness is also evaluated. Take Scenario 1 as an example, Fig.~\ref{fig:tt0} shows the data completeness rate under different timeout $T_{T0}$ values. Given a specific timeout $T_{T0}$ value, the data completeness rate of a scenario with QoS control outperforms the scenario without QoS control. In another word, the data marking mechanism at the PMU fog can effectively take more advantage of the QoS control mechanism in the communication network. Such that a lower $T_{T0}$ can be applied, which further reduces the ETE delay with a guaranteed data completeness rate.  
\section{Conclusions}
\label{sec:conclusions}
The efficiency of the PMU communication network is important for latency strict WAMC applications. In this paper, a prioritized data transmission mechanism for PMU measurement data flow is proposed. Inspired by the edge-fog-cloud computing paradigm, the measurement data of fault or disturbance is identified at the PMUs and marked with higher priority, such that the QoS mechanism existing in the communication networks can be leveraged to selectively accelerate the data delivery. For this purpose, two anomaly detection methods are studied at the PMU fog layer, SSA and KNN. Experimental results prove that both methods are fit at the fog layer with different advantages and constraints. The simulation of the communication network shows the prioritized transmission mechanism can effectively reduce the data flow ETE delay without sacrificing data completeness.